\documentclass[aip,apl,reprint,twoside,floatfix]{revtex4-1} 
\pdfoutput=1
\usepackage{graphicx}
\usepackage{amsmath}

\draft 

\usepackage{bm}

\begin{document}


\title{Magnetic power inverter: AC voltage generation from DC magnetic fields} 



\author{Jun'ichi Ieda}
\affiliation{%
Advanced Science Research Center, Japan Atomic Energy Agency, Tokai 319-1195, Japan}
\affiliation{%
CREST, Japan Science and Technology Agency, Tokyo 102-0075, Japan}
\author{Sadamichi Maekawa }
\affiliation{%
Advanced Science Research Center, Japan Atomic Energy Agency, Tokai 319-1195, Japan}
\affiliation{%
CREST, Japan Science and Technology Agency, Tokyo 102-0075, Japan}


\date{\today}

\begin{abstract}
We propose a method that allows power conversion from DC magnetic fields to AC electric voltages using domain wall (DW) motion in ferromagnetic nanowires. The device concept relies on spinmotive force, voltage generation due to magnetization dynamics. Sinusoidal modulation of the nanowire width introduces a periodic potential for a DW the gradient of which exerts variable pressure on the traveling DW. This results in time variation of the DW precession frequency and the associated voltage. Using a one-dimensional model we show that the frequency and amplitude of the AC outputs can be tuned by the DC magnetic fields and wire-design.
\end{abstract}


\maketitle 

%
%

Spinmotive force is one of the emerging concepts in spintronics\cite{CSE06, SC12}, which allows direct conversion of magnetic energy to electric voltage in magnetic nanostructures.
The embodiment of this effect was considered in a ferromagnetic nanowire containing a magnetic domain wall (DW)\cite{BarMae07}.
Application of a DC magnetic field induces DW motion along the nanowire.
The exchange interaction mediates energy transfer between the local moment and the conduction electron spin.
As a result, a DC electrical voltage develops across the DW where the source of the electric power is the Zeeman energy.
The rate of conversion is given by $P\hbar\gamma/e \sim 10^2$ $\mu$V/T  where $P$ is the spin polarization of the ferromagnetic material, $\hbar$ is the Planck constant divided by $2\pi$, $\gamma$ is the gyromagnetic ratio, and $e$ is the elementary charge.
Thanks to advances in microfabrication technology, the prediction has been confirmed in controlled experiments using the lock-in technique\cite{Yang09} and the time-domain measurement\cite{HayIed12}. 

Striking features of the spinmotive force are listed as follows: 
1) In contrast to the inductive voltage where the time variation of a magnetic flux is required, solely a DC magnetic field can generate an electric voltage (see, e.g., Fig.2a in Hai \emph{et al.} \cite{Hai09}). 
2) The conversion rate is represented by fundamental constants apart from the material-dependent $P$, offering high energy conversion efficiency and precise determination of important parameter $P$ by measuring the output voltages as a function of applied fields.
3) Applications using this effect can operate as active devices with zero stand-by power\cite{BarIedMae06} and such a power-conversion ability between magnetic and electric systems might open up spin-based power electronics. 

Recently, in addition to the DC voltage generation from the field-induced DW motion, gyration of a magnetic vortex core has been shown to generate AC voltages from applied AC magnetic fields (microwaves) where contribution of the standard inductive voltage is carefully separated from the spinmotive force signal\cite{Tanabe12}.
Moreover, the AC to DC magnetic power converter (i.e., from AC magnetic fields to DC voltages) was demonstrated in a comb-shaped ferromagnetic thin film in which spatially selective ferromagnetic resonance is excited\cite{YamSas11}. 
However, the inversion of the latter device, i.e., a \emph{magnetic power inverter}, has not been established yet.

\begin{figure}[b]
\begin{center}
\includegraphics[width=85mm]{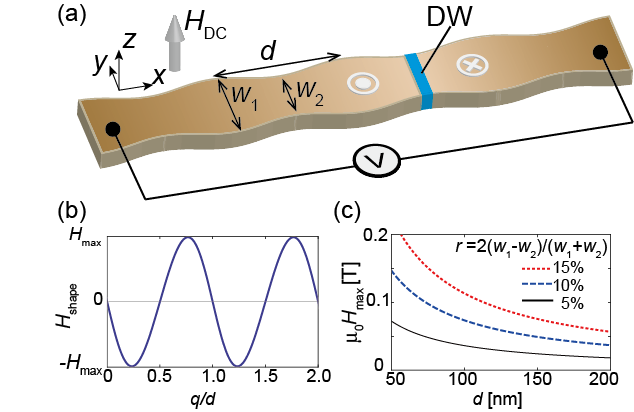}
\end{center}
        \caption{ (a) The schematic illustration of a periodically modulated nanowire with perpendicular magnetization containing a DW.
                      The wire-width $w(x)$ ($w_1\ge w \ge w_2$) is modulated sinusoidally with the period $d$.
                  (b) The shape-effect magnetic field, $H_\mathrm{shape}$ calculated by Eqs.~(\ref{shape}) and (\ref{h})
                  as a function of the DW position $q$.
                  (c) The peak-hight of the shape-effect magnetic field $H_\mathrm{max}$ as a function of the modulation period with $r=5$,
                  $10$, $15\%$ where typical parameters for a Co/Ni multilayer are used. 
                }
\label{fig1}
\end{figure}

In this paper, we propose a method of AC voltage generation due to application of a DC magnetic field using DW motion in a periodically modulated nanowire as shown in Fig.~\ref{fig1}(a). 
Since a DW has a surface energy due to the exchange interaction and magnetic anisotropy
the modulation of the wire width introduces a variation of the potential energy for the DW.
The geometrical confinement arising from this effect is commonly used for pinning DWs\cite{Parkin08,Klaui08}.
It has been suggested that DW motion can be induced solely by the shape effect
and the intrinsic magnetic energy of the DW can be exploited via the spinmotive force mechanism\cite{BarIedMae06,shape}.


%
%


We consider a perpendicularly magnetized thin wire, 
 in which the thickness of the wire is constant while the width is modulated as
\begin{equation}
w(x)  =  \bar{w}\left[ 1-2r \cos (2\pi x/d) \right],
\label{shape}\end{equation}
with the average width $\bar{w}=(w_1+w_2)/2$, modulation rate $r=(w_1-w_2)/\bar{w}$,
and period $d$ as indicated in Fig.~\ref{fig1}(a).
A DW is prepared in the nanowire and a DC magnetic field $H_\mathrm{DC}$ is applied 
perpendicular to the plane (the $z$ direction) to drive the DW along the wire (the $x$ direction).
Then the voltage induced by the DW motion is measured between the ends of the wire.

First, we analytically investigate influence of the width modulation on the spinmotive force using a collective-coordinate model in which the dynamics of a DW is characterized by two collective coordinates, the center position $q$ and the tilt angle $\psi$ of the DW plane relative to the easy-plane.
Importantly, in addition to the applied field, a DW in the spatially nonuniform nanowire is subjected to the ``shape-effect field'' due to the variation of the surface energy\cite{BarIedMae06,IedSugMae10}:
\begin{equation}
H_\mathrm{shape}=\mp\frac{\sigma}{2M_\mathrm{s}}\left. \frac{d}{d x}\ln w(x)\right|_{x=q},
\label{h}\end{equation}
where $\sigma=\sigma_0\left(1+Q^{-1}\sin^2\psi \right)^{1/2}$ is the surface energy density of a DW and $M_\mathrm{s}$ is the saturation magnetization.
Here $\sigma_0=4\sqrt{A_\mathrm{s}K_\mathrm{u}}$ with $A_\mathrm{s}$ the exchange stiffness constant, $K_\mathrm{u}$ the uniaxial anisotropy constant, and $Q$ is the ratio of $K_\mathrm{u}$ to the hard-axis anisotropy constant. 
Here and hereafter the upper (lower) sign corresponds to the up-down (down-up) DW for the present coordinate-system.
Figure \ref{fig1}(b) illustrates $H_\mathrm{shape}$ as a function of the DW position for the sinusoidally modulated width~(\ref{shape}).
The peak hight $H_\mathrm{max}$ is given by 
\begin{equation}
H_\mathrm{max}=\frac{2\pi\sigma}{M_\mathrm{s}}\frac{r}{d\sqrt{1-4r^2}}.
\label{hmax}
\end{equation}
In Fig.~\ref{fig1}(c), we show the shape dependence of Eq.~(\ref{hmax}).
For $|H_\mathrm{DC}|\lesssim H_\mathrm{max}$, a DW is trapped by the potential well and the total field acting on the DW becomes zero.
This leads to the threshold field for the voltage generation.
When a DW propagates along the sinusoidal wire the AC voltage with the amplitude $(P\hbar\gamma/e)\mu_0H_\mathrm{max}$ is expected.
In the uniform limit, $r\to0$ or $d\to\infty$, the shape effect disappears.

In the one-dimension model, the voltage due to DW dynamics is given by\cite{BarMae07}
\begin{equation}
V  =  \pm \frac{P\hbar}{e} \dot{\psi}.
\label{v}\end{equation}
Note that the growing side of the magnetic domain becomes high voltage being independent of the wall type\cite{HayIed12}.
To evaluate this voltage, the DW dynamics has to be calculated.
The time evolution of ($q,\psi$) is described by the reduced Landau-Lifshiz-Gilbert equations including the shape-field,
\begin{align}
\dot{q}     =&  \pm\frac{ \Delta \gamma\mu_0 } { 1 + \alpha^2 }  \left[  \alpha \left( H_\mathrm{DC} + H_\mathrm{shape} \right)  +         \frac {H_K} {2} \sin 2 \psi   \right],    \label{q}\\
\dot{\psi}  =&  \frac{ \gamma\mu_0 }        { 1 + \alpha^2 }  \left( H_\mathrm{DC} + H_\mathrm{shape}          -  \alpha \frac {H_K} {2} \sin 2 \psi   \right),    \label{psi}
\end{align}
where $\alpha$ is the Gilbert damping constant, $\Delta=\sqrt{A_\mathrm{s}/K_\mathrm{u}}$ is the wall width, $H_K$ is the hard-axis anisotropy field, and $\mu_0$ is the magnetic constant. 
In a uniform nanowire the time-average DW velocity $\bar{v}_\mathrm{DW} $
is well described by $\bar{v}_\mathrm{DW} = \alpha\Delta \gamma\mu_0 H_\mathrm{DC}/(1 + \alpha^2)$
for $|H_\mathrm{DC}|> H_\mathrm{W} $ where $H_\mathrm{W}=\alpha H_K/2$ is the Walker breakdown field.
In the sinusoidal nanowire, the condition for the Walker breakdown is replaced by $|H_\mathrm{DC}+H_\mathrm{shape}|> H_\mathrm{W} $.
The one-dimensional description is more accurate for perpendicularly magnetized nanowires with high magnetic anisotropy such as Co/Ni multilayers\cite{CoNiparam1} than for Permalloy nanowires where the DW structure is deformed during the propagation\cite{Hayashi10}. 
It has been shown that the rigid DW motion supported by the high magnetic anisotropy is favorable for stable generation of the spinmotive force\cite{YamIedMae12}.

In the following, we numerically solve the one-dimensional model (\ref{q}) and (\ref{psi}) for the sinusoidal nanowire defined by (\ref{h}) to clarify the AC output characteristics of the DW-induced voltages.
%
%

\begin{figure}[t]
\begin{center}
\includegraphics[width=85mm]{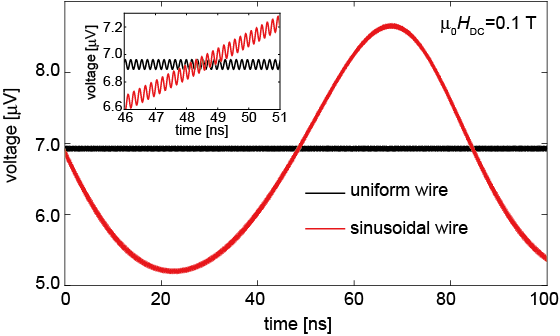}
\end{center}
        \caption{The time evolution of 
        the output voltage signal 
        for a uniform wire (black) and a sinusoidal wire (red) with $d=150$ nm and $r=5$ \%. The same DC magnetic field, $\mu_0H_\mathrm{DC}=0.1$ T, is applied for both geometries. 
        The inset is the enlarged view showing the oscillation due to the anisotropy field.
                }
\label{fig2}
\end{figure}

Figure \ref{fig2} shows the calculated 
voltage signals due to DW motion in the sinusoidal, as well as the uniform (reference), nanowires.
We employ typical values for a Co/Ni multilayer; $\gamma=1.76\times 10^{11}$ s${}^{-1}\cdot$T${}^{-1}$, $\alpha=0.02$, $P=0.6$, $M_\mathrm{s}=0.85$ T, $A_\mathrm{s}=1.3\times 10^{-11}$ J/m, $K_\mathrm{u}=4.0\times 10^5$ J/m$^3$, and $\mu_0H_K=50$ mT. 
We set the modulation parameters for the sinusoidal wire as $d=150$ nm, $r=5$ \%, and the applied DC magnetic field $\mu_0H_\mathrm{DC}=0.1$ T.
In addition to the small-amplitude oscillation associated to the Walker breakdown as shown in the inset of Fig.~\ref{fig2}, 
the AC component with the amplitude $\sim\pm1.7$ $\mu$V and the period $\sim 85$ ns appears for the sinusoidal nanowire. 
The base line is the DC component $(P\hbar\gamma/e)\mu_0H_\mathrm{DC}=6.93$ $\mu$V.
The asymmetry found in positive and negative waveforms of the AC component is caused by the difference of the DW velocity, $\dot{q}$, for the shape-field assisting and being against the DW propagation respectively.
We discuss the shape dependence of the asymmetry later.

\begin{figure}[t]
\begin{center}
\includegraphics[width=85mm]{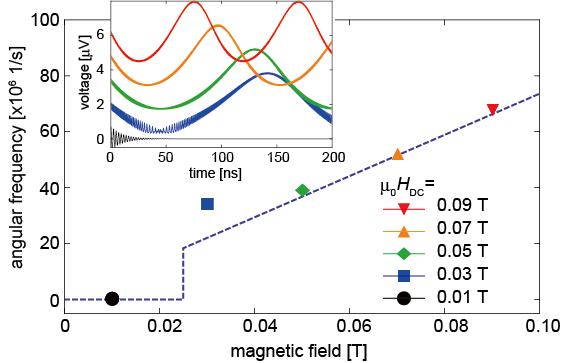}
\end{center}
        \caption{The applied magnetic field dependence of the output voltage signal frequency.  The symbols represent the second maximum Fourier component of the representative voltage signals, shown in the inset, for $\mu_0H_\mathrm{DC}=0.01$--0.09 T as indicated in the legend. The sinusoidal wire parameters are $d=150$ nm and $r=5$ \%. The dashed curve shows the analytical formula (\ref{omega}). 
                }
\label{fig3}
\end{figure}

The AC voltage amplitude agrees well with the predicted value, $(P\hbar\gamma/e)\mu_0H_\mathrm{max}=1.72$ $\mu$V, using Eq.~(\ref{hmax}) for the present geometry, which is basically independent of $H_\mathrm{DC}$ as shown in the inset of Fig.~\ref{fig3}.
The period $T$ is determined by $T  =  \int_0^d d q/\dot{q} \simeq d/\bar{v}_\mathrm{DW}$. 


To see the $H_\mathrm{DC}$ dependence of the AC component, we perform the Fourier analysis of the calculated voltage signals. 
We focus on the angular frequency $\omega_\mathrm{AC}=2\pi/T$ of the AC component induced by the shape effect.
In the frequency domain, we identify the second highest peak as the shape-induced frequency as well as the highest DC component and the small peak of the anisotropy origin (the Walker breakdown).

In Fig.~\ref{fig3}, we plot $\omega_\mathrm{AC}$ for representative magnetic fields, $\mu_0H_\mathrm{DC}=$ 10, 30, 50, 70, and 90 mT. 
The parameters used here are the same as in Fig.~\ref{fig2}. 
The general features of the frequency can be understood by the approximate expression:
\begin{align}
\omega_\mathrm{AC}  \simeq  
\left\{
\begin{array}{cl}
\displaystyle
\frac{ 2\pi \alpha\Delta \gamma\mu_0 }{ (1 + \alpha^2)d }H_\mathrm{DC} & (|H_\mathrm{DC}| \ge H_\mathrm{max}),\\
0 & (|H_\mathrm{DC}|<H_\mathrm{max}),
\end{array} \right. 
\label{omega}\end{align}
as indicated by the dashed line in Fig.~\ref{fig3}.
The deviation from Eq.~(\ref{omega}) near the threshold value can be attributed to the nonlinear DW dynamics\cite{IedSugMae10} where the contribution from the anisotropy field plays a decisive role. 
It is interesting to investigate such nonlinear behaviors in more detail but we leave these problems for future work. 

In the inset of Fig.~\ref{fig3}, we show the time-domain signals for corresponding $H_\mathrm{DC}$. 
Note that the threshold value is $\mu_0H_\mathrm{max}\simeq 25$ mT for the present geometry.
For $\mu_0H_\mathrm{DC}=10$ mT, the voltage signal becomes zero as the DW is trapped by the geometrical potential. 
For $\mu_0H_\mathrm{DC}=30$ mT, the asymmetry of the AC signal is pronounced. 
On the other hand, for the larger $H_\mathrm{DC}$ the monochromaticity increases since the relative variation of the DW velocity due to the shape-field becomes small. 

%
%

\begin{figure}[t]
\begin{center}
\includegraphics[width=85mm]{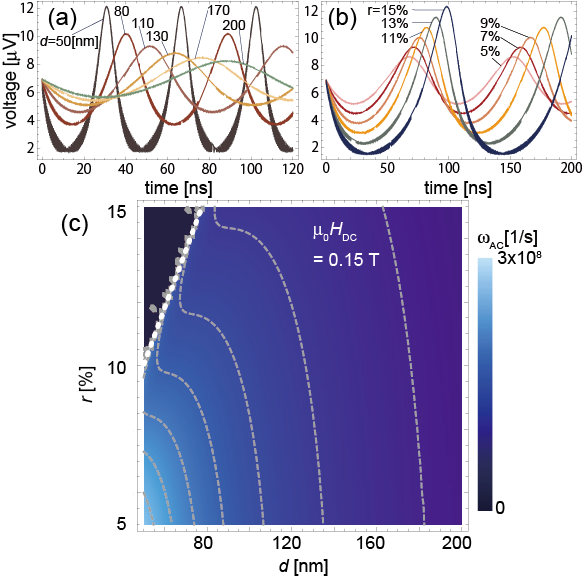}
\end{center}
        \caption{ The wire-geometry dependence of the output voltage signals. 
	        (a) The period $d$ of the nanowire is varied from 50--200 nm with $\mu_0H_\mathrm{DC}=0.1$ T and $r=5$ \% being fixed.
	        (b) The wire-modulation ratio $r$ is varied from 5--15 \% for constant $\mu_0H_\mathrm{DC}=0.1$ T and $d=100$ nm. 
	        (c) The angular frequency of the AC component $\omega_\mathrm{AC}$ as a function of $d$ and $r$ 
	        for $\mu_0H_\mathrm{DC}=0.15$ T.
	        The dotted line indicates the threshold for the DW deppining. 
                } 
\label{fig4}
\end{figure}

Next we investigate the geometry dependence of the AC component in Co/Ni wires.
In Figs.~\ref{fig4}(a) and \ref{fig4}(b), we show the time-domain voltage signals for $\mu_0H_\mathrm{DC}=0.1$ T varying $d$ (with $r=5$ \% fixed) and $r$ (with $d=150$ nm fixed) of sinusoidal nanowires respectively.
While the AC amplitudes are proportional to $1/d$ and $\simeq r$ as estimated by Eq.~(\ref{hmax}), the AC frequencies diminish with increasing both $d$ and $r$.
These features will be useful for engineering the device characteristics, AC amplitude and frequency, independently.
Figure \ref{fig4} (c) summaries the shape dependence of the AC frequency for  $\mu_0H_\mathrm{DC}=0.15$ T.
For the area $H_\mathrm{max}\gtrsim H_\mathrm{DC}$, DWs are trapped and the voltage is not generated.

Finally, we remark that the periodic potential for a DW can be realized by modulating not only the width but the thickness of a nanowire, or the radius of a magnetic nanotube. 
In a cylindrical nanotubes, the hard-axis anisotropy is absent, leading to fascinating features for DW dynamics\cite{Hertel11}, and thus, for the associated voltages.
Different materials have different AC characteristics, \emph{e.g.}, the faster DW motion could achieve the GHz operation.
This also merits further investigation.

%
%

In summary, we have presented the concept of a magnetic power inverter: power conversion from DC magnetic field to AC electric voltage.
The device consists of a periodically modulated ferromagnetic nanowire with a DW.
We have investigated systematically the output voltage characteristics as a function of the input DC magnetic fields and the geometrical parameters of the nanowire using the one-dimensional model for the DW.
We have shown that by tuning the magnetic field and the wire geometry the variable frequency ranging from MHz to GHz can be achieved.
The proposed device operates as an active element in future spin-based power electronics, or \emph{power spintronics}.

%
%

We are grateful to Y. Yamane for fruitful discussions on this work.
This research was supported by MEXT KAKENHI Grant Number 24740247.

%
%


\begin{thebibliography}{99}
\bibitem{CSE06} S. Maekawa ed., {\em Concepts in Spin Electronics} (Oxford University Press, Oxford, 2006).
\bibitem{SC12} S. Maekawa, S. O. Valenzuela, E. Saitoh, and T. Kimura eds., {\em Spin Current} (Oxford University Press, Oxford, 2012).
\bibitem{BarMae07} S. E. Barnes and S. Maekawa, Phys. Rev. Lett. {\bf 98}, 246601 (2007).
\bibitem{Yang09} S. A. Yang G. S. D. Beach, C. Knutson, D. Xiao, Q. Niu, M. Tsoi, and J. L. Erskine, Phys. Rev. Lett. {\bf 102}, 067201 (2009). 
\bibitem{HayIed12} M. Hayashi, J. Ieda, Y. Yamane, J. Ohe, Y. K. Takahashi, S. Mitani, and S. Maekawa, Phys. Rev. Lett. {\bf 108}, 147202 (2012).
\bibitem{Hai09} P. N. Hai, S. Ohya, M. Tanaka, S. E. Barnes, and S. Maekawa, Nature {\bf 458}, 489 (2009).
\bibitem{BarIedMae06} S. E. Barnes, J. Ieda, and S. Maekawa, Appl. Phys. Lett. {\bf 89}, 122507 (2006).
\bibitem{Tanabe12} K. Tanabe, D. Chiba, J. Ohe, S. Kasai, H. Kohno, S. E. Barnes, S. Maekawa, K. Kobayashi, and T. Ono, Nature Commun. {\bf 3}, 845 (2012).
\bibitem{YamSas11} Y. Yamane, K. Sasage, A. Toshu, K. Harii, J. Ohe, J. Ieda, S. E. Barnes, E. Saitoh, and S. Maekawa, Phys. Rev. Lett. {\bf{107}}, 236602 (2011).
\bibitem{Parkin08} S. S. P. Parkin, M. Hayashi, and L. Thomas, Science {\bf 320}, 190 (2008).
\bibitem{Klaui08} M. Kl\"aui, J. Phys.: Condens. Matter {\bf 20}, 313001 (2008).
\bibitem{shape}    Y. Yamane, J. Ieda, J. Ohe, S. E. Barnes, and S. Maekawa, Appl. Phys. Exp. {\bf{4}}, 093003 (2011).
\bibitem{IedSugMae10}     J. Ieda, H. Sugishita, and S. Maekawa, J. Magn. Magn. Mat. {\bf{322}}, 1363 (2010).
\bibitem{CoNiparam1}   T. Koyama, D. Chiba, K. Ueda, H. Tanigawa, S. Fukami, T. Suzuki, N. Ohshima, N. Ishiwata, Y. Nakatani, and T. Ono, Appl. Phys. Lett. {\bf{98}}, 192509 (2011).
\bibitem{Hayashi10}  M. Hayashi, S. Kasai, and S. Mitani, Appl. Phys. Exp. {\bf{3}}, 113004 (2010).
\bibitem{YamIedMae12} Y. Yamane, J. Ieda, and S. Maekawa, Appl. Phys. Lett. {\bf{100}}, 162401 (2012). 
\bibitem{Hertel11} M. Yan, C. Andreas, A. K\'{a}kay, F. Garc\'{i}a-S\'{a}nchez, and R. Hertel, Appl. Phys. Lett. {\bf{99}}, 12505 (2011).

\end{thebibliography}
\end{document}